\documentclass[article,
showpacs,superscriptaddress,
floatfix,
preprint]{revtex4-1}

\usepackage{graphicx}
\usepackage{color}
\usepackage{siunitx}
\usepackage{upgreek}
\usepackage{epstopdf}
\usepackage{braket}
\usepackage{amssymb,amsmath}

\begin{document}
	
	\title{Observation of electromagnetically induced absorption in a vee + ladder system}
	
	\author{Mangesh Bhattarai}
	\affiliation{Department of Physics, Indian Institute of Science, Bangalore - 560012}
	\author{Vineet Bharti}
	\email{Currently at Institute for Molecular Science, National Institutes of Natural Sciences, Okazaki, Aichi 444-8585, Japan}
	\affiliation{Department of Physics, Indian Institute of Science, Bangalore - 560012}
	\author{Sambit Banerjee}
	\email{Currently at Department of Physics and Astronomy, Purdue University, West Lafayette, Indiana 47907, USA}
	\affiliation{Department of Physics, Indian Institute of Science, Bangalore - 560012}
	\author{Vasant Natarajan}
	\email{vasant@physics.iisc.ernet.in}
	\affiliation{Department of Physics, Indian Institute of Science, Bangalore - 560012}
	
	\begin{abstract}
		We experimentally demonstrate electromagnetically induced absorption (EIA) in a vee + ladder system. The experiment is done using the low-lying energy levels of $^{87}$Rb. A theoretical model of the system is made that reproduces the experimental results. We study the dependence of the characteristics of the EIA resonance on various combinations of the different control powers. We also explore the contribution of various incoherent phenomena that affect the EIA signal.
	\end{abstract}
	
	\maketitle
	
	\section{Introduction}
	The phenomenon of EIT has been well studied over the years in lambda ($\Lambda$), vee (V) and ladder ($\Xi$) type systems. EIT creates transparency for a weak probe in an otherwise absorbing medium in the presence of a strong control beam by nonlinear modification of the susceptibility of the medium \cite{FIM2005}. This has led to a variety of applications including in slow light \cite{HHD1999,BKR1999,KSZ1999}, storage of light \cite{LDB2001,PFM2001}, lasing without inversion \cite{ZLN1995}, four wave mixing \cite{HKD1995}, giant Kerr effect \cite{SCI1996,MBB2008}, high resolution spectroscopy \cite{KPW2005}, laser frequency stabilization \cite{MLK2004}, and phase stabilization of independent sources \cite{BKN2019}.
	
	Similar to the phenomenon of EIT is the phenomenon of electromagnetically induced absorption (EIA), except that the probe shows an enhanced absorption  and a negative dispersion slope (as opposed to the positive slope in EIT) at line center. EIA has been reported in four level N type systems \cite{YZR2002,GWR2004,BMW2009} with two control beams and one probe beam, in degenerate two level systems \cite{CSB2011,DBD2016,LBA1999} and recently in a three level system \cite{WBK2015}. Recently, we have theoretically proposed a new kind of four-level EIA system in a vee + ladder configuration \cite{BHN2015}. Using this scheme, we have presented the first theoretical observations of Rydberg EIA \cite{BWN2016} and tuning of the group velocity of light between sub- and super-luminal propagation using a Rydberg state \cite{BHN2017}. This system has drawn much attention, and has been utilized for the investigation of amplitude noise in non-classical states of light \cite{MLY2016}, optical bistability \cite{HSK2017}, generation of an all-optical diffraction grating \cite{BOS2018} and atom localization \cite{HSK2018}. In this work, we report the first experimental observation of EIA in such a system. 
	
		\begin{figure}
			\centering
			\includegraphics[width=0.85\textwidth]{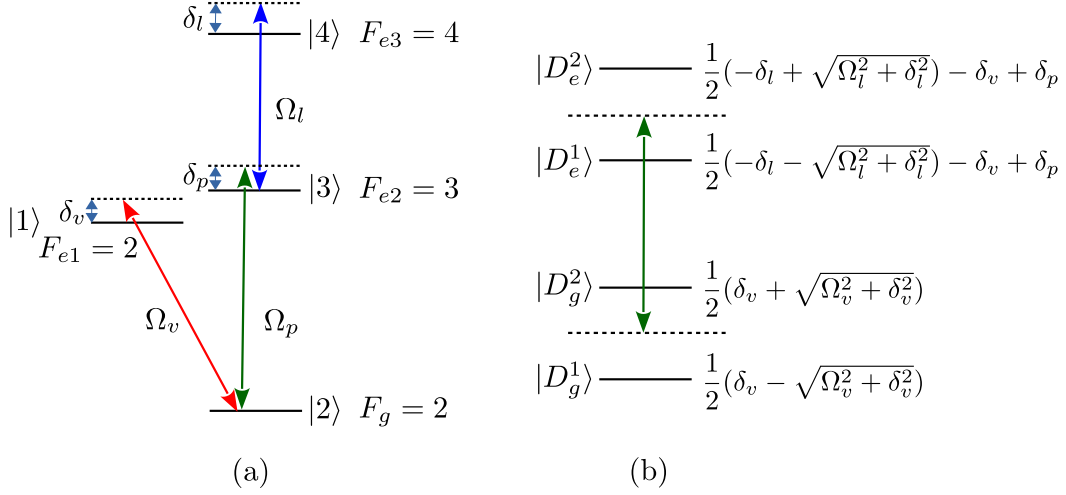}
			\caption{(a) Energy levels depicting the formation of the vee + ladder system, (b) 	Levels represented in a dressed state picture.}
			\label{fig:levels}
		\end{figure}

	We begin with a qualitative look at the phenomenon of EIA in terms of the dressed state picture. The vee + ladder system is formed by $^{87}$Rb interacting with two copropagating fields $\Omega_v$ and $\Omega_p$ and a counterpropagating field $\Omega_l$, coupling the levels shown in Fig.~\ref{fig:levels} with detunings $\delta_v$, $\delta_p$ and $\delta_l$, respectively. The interaction Hamiltonian is given as \cite{BHN2015}
	\begin{equation}
	\begin{aligned} 
	H =  &-\frac{\hbar}{2}\left[\Omega_v \ket{1}\bra{2}+\Omega_p \ket{2}\bra{3}+\Omega_l \ket{3}\bra{4}\right]+\rm h.c. \\
	&-\hbar \left[0 \ket{1}\bra{1} -\delta_v \ket{2}\bra{2}+(-\delta_v+\delta_p) \ket{3}\bra{3}+(-\delta_v+\delta_p+\delta_l) \ket{4}\bra{4} \right] 
	\end{aligned}
	\end{equation}
 The field $\Omega_v$  mixes the states  $\ket{1}, \ket{2}$ to form the dressed states $\ket{D_g^1}, \ket{D_g^2}$ and  the field $\Omega_l$ mixes $\ket{3}, \ket{4}$ creating the dressed states $\ket{D_e^1}, \ket{D_e^2}$. The dressed state energy is as indicated in Fig.~\ref{fig:levels}(b). A weak probe $\Omega_p$ scanned across resonance gives 4 absorption peaks at the following values of probe detuning, $\delta_p = $
	 \begin{equation}
	 \begin{aligned}
	 \centering
	 & 1/2(-\delta_l+\delta_v+\sqrt{\delta_l^2+\Omega_l^2}-\sqrt{\delta_v^2+\Omega_v^2}),\\
	 & 1/2(-\delta_l+\delta_v-\sqrt{\delta_l^2+\Omega_l^2}+\sqrt{\delta_v^2+\Omega_v^2}),\\
	 & 1/2(-\delta_l+\delta_v-\sqrt{\delta_l^2+\Omega_l^2}-\sqrt{\delta_v^2+\Omega_v^2}),\\
	 & 1/2(-\delta_l+\delta_v+\sqrt{\delta_l^2+\Omega_l^2}+\sqrt{\delta_v^2+\Omega_v^2})
	 \end{aligned}
	 \end{equation}
	 For an atom moving with a velocity $v$, the position of the peaks are found by the substitution $\delta_p \to \delta_p -k v$, $\delta_v \to \delta_v -k v$, $\delta_l \to \delta_l +k v$. Along with the condition $\Omega_v=\Omega_l$ and $\delta_v=-\delta_l$, the peaks lie at $\delta_p = \delta_v, \delta_v + \sqrt{\Omega_v^2+ (\delta_v-k v)^2},\delta_v - \sqrt{\Omega_v^2+ (\delta_v-k v)^2} $.

	 The first peak has no dependence on the velocity of atom, and the second and the third peaks lie symmetrically on the right and left side of the first peak, respectively, for any $v$. In a thermal ensemble, the first peak for all velocity groups occurs at the same value of the detuning and consequently grows giving a narrow absorption feature. The peaks on either side get spread around different values of the detunings, only to give a small rise to the absorption beyond detuning larger than $\Omega_v$ from the first peak. Thus the resonance condition for EIA can be stated as 
	 \begin{equation}
	 \delta_p -\delta_v =0 \quad {\text{and}} \quad \delta_p + \delta_l=0
	 \end{equation}
	 which is just a statement about the simultaneous EIT condition for the vee and ladder branches of the system. When both the control fields are on resonance ($\delta_v=\delta_l = 0$), EIA occurs at $\delta_p = 0$. 
	
		\begin{figure}
			\centering
			\includegraphics[width=0.8\textwidth]{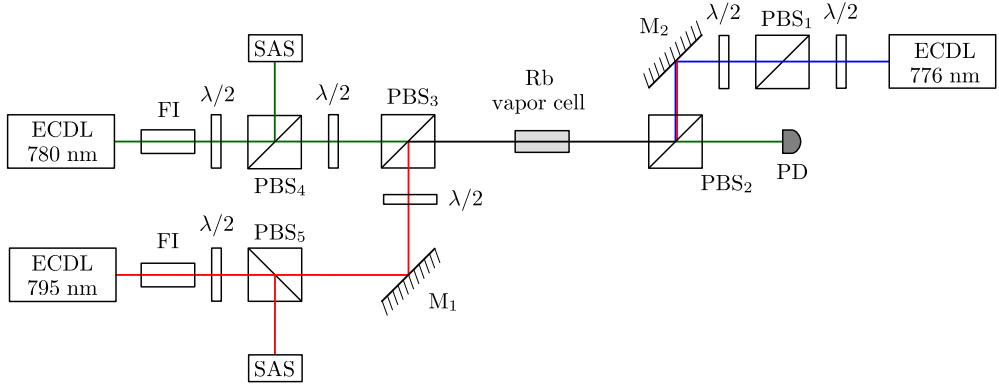}
			\caption{Experimental schematic. Figure key: FI -- Faraday isolator; SAS -- saturated absorption spectroscopy; PBS -- polarizing beam splitter cube; PD -- photo-diode; $\lambda/2$ -- half wave retardation plate; M -- dielectric mirror.}
			\label{fig:my_label}
		\end{figure}
	
	\section{Experimental details}
	The experimental setup is shown in Fig.~\ref{fig:my_label}. It consists of three feedback-stabilized ECDLs --- two home-built lasers operating around 780 nm and 795 nm and a Toptica DL Pro laser operating around 776 nm. The beams from the lasers are Gaussian and have $1/e^2$ diameters of 1.4 mm $\times$ 2.2 mm, 1.9 mm $\times$ 1.6 mm and 1.9 mm $\times$ 1.6 mm, respectively. The 780 nm laser is operated around the $F_g=2 \to F_{e2}=3$ transition in the D$_2$ line of $^{87}$Rb. A portion of the laser goes to a saturated absorption spectrometer used as frequency reference and the rest to the main experiment. The 795 nm laser on the D$_1$ line of $^{87}$Rb is locked to the $F_g=2 \to F_{e1}=2$ transition by modulation of the injection current using another saturated absorption spectroscopy setup. The 780 nm laser serves as the probe and the 795 nm laser (henceforth referred to as the vee-control) serves as the control for the vee-type EIT branch of the total setup. These two beams are combined at PBS$_3$ and co-propagated to a Rb vapor cell. The vapor cell is cylindrical in shape, of length 50 mm and diameter 25 mm. It contains the 2 stable Rb isotopes in their natural abundances (72.17 \% $^{85}$Rb and 27.83 \% $^{87}$Rb). It is placed inside a framework holding three pairs of Helmholtz coils which are used to null the magnetic field around the cell. Counter propagating to the probe and the vee-control beams within the Rb vapor cell is the 776 nm laser beam coupling the $F_{e2}=3 \to 5D_{5/2}$ line in $^{87}$Rb. This beam forms the control beam for the ladder EIT branch of the system and will be reffered to as ladder-control. The PBS$_2$ after the vapor cell serves to counter-propagate the ladder-control beam as well as separate the probe beam from the vee-control beam. The probe is collected on the photo-diode PD and the spectrum is recorded on a digital storage oscilloscope.
		
	\section{Results and discussion}
\begin{figure}
		\centering
		\includegraphics[width=0.65\textwidth]{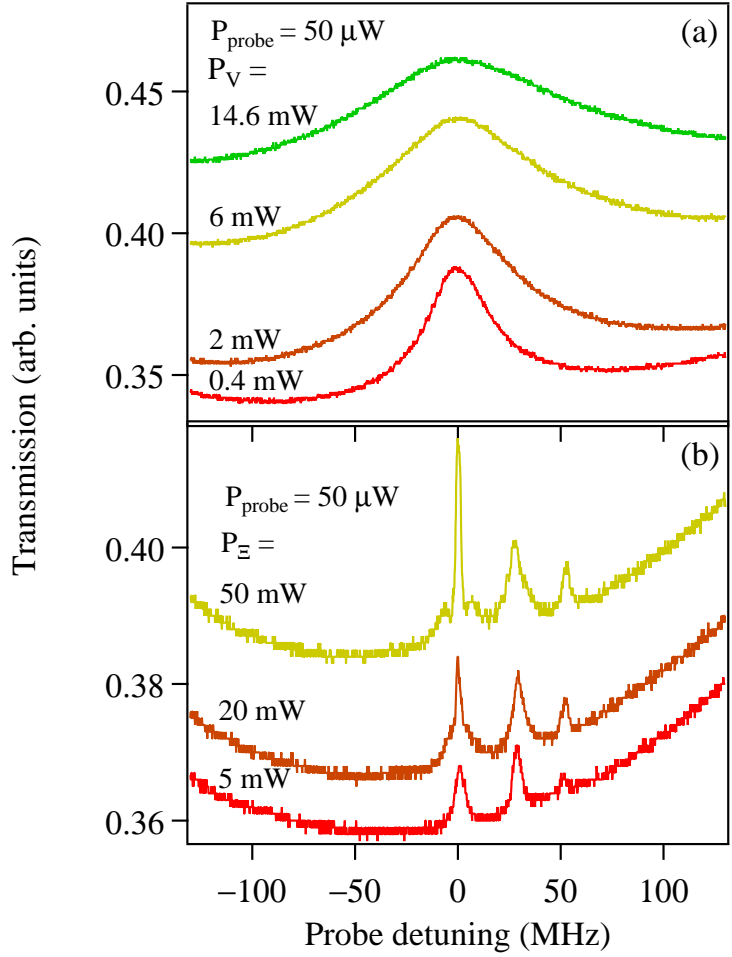}
		\caption{EIT for different systems. (a) Vee (b) Ladder}
		\label{fig:EIT}
	\end{figure}

	We begin by summarizing EITs in vee and ladder systems and demonstrate transformation of EIT to EIA when both controls are simultaneously operating. Figs.~\ref{fig:EIT}(a) and (b) show EITs in vee and ladder systems, respectively, for different powers in the respective control beams.
	
	The vee system is formed by the probe on $F_g=2 \to F_{e2}=3$ in the D$_2$ line and the control on $F_g=2 \to F_{e1}=2$ in the D$_1$ line. As has been studied earlier \cite{FSM1995}, transparency in the vee system is contributed to by a lot of other factors besides coherent EIT effects, namely (i) saturation effect which is ever-present as the control is coupled to the ground state, (ii) optical pumping to the uncoupled hyperfine level and (iii) optical pumping within the magnetic sublevels. In the current experiment, optical pumping to the uncoupled hyperfine level exists because the control couples the $F_g=2$ ground state to $F_{e1}=2$ excited state which could decay to the ground state $F_g = 1$ not coupled by any light. This adds to probe transparency by loss of atoms from the interacting levels. 
	
	It is seen in Fig.~\ref{fig:EIT}(a) that smaller values of the control power (increase from 0.4 mW to 2 mW) increases the EIT height. Larger control powers (6 mW and 14.6 mW), however, seriously depletes the number of atoms available for EIT decreasing its strength. A strong control addresses transition for a larger group of velocities which increases the background transmission by increasing the probe transparency on either side of the resonance. This process at the same time decreases the strength and increases the width of the EIT. The broadening effect is also associated with the increase in the transparency window caused by higher Rabi frequency of the control.  The other effect contributing to transparency is optical pumping within the magnetic sublevels. It depends on the polarization combination of the probe and the control beams. Having orthogonally linearly polarized control and probe adds to transparency of the probe by Zeeman optical pumping.
	
	Fig.~\ref{fig:EIT}(b) shows EIT formed in the ladder system comprised of the probe on the $F_g=2 \to F_{e2}=3$ at the D$_2$ line and the control on the $F_{e2}=3 \to F_{e3}=4,3,2$ in the hyperfine levels of 5D$_{5/2}$ state. In the figure, the transmission peak at zero detuning of the probe corresponds to the EIT with the control on the $F_{e2}=3 \to F_{e3}=4$ transition. The other two peaks at the positive value of detunings are EITs corresponding to $F_{e3}=3,2$ levels. The different powers of control used depicts the transformation of lineshape of the EIT signal. The EIT corresponding to $F_{e3}=4$ transition distinctively shows combination of two spectral profiles for powers of 20 mW and 50 mW. The broad profile is caused by saturation effect on the probe transition and optical pumping effects due to some population transfer induced by the probe beam. This has been reported earlier in literature as double resonance optical pumping (DROP) effect \cite{MLK2008}. Over this profile a narrow EIT feature is formed. The strength of the EIT signal increases with the control power, as is evident from the figure. For the case of 5 mW power in the control beam, there is no clear distinction between the DROP effect and EIT. A similar undifferentiation of the different spectral profiles are seen in EITs for $F_{e3}=3, 2$ levels. Atoms in these levels can decay to the uncoupled $F_g=1$ hyperfine level in the ground state via the intermediate  $F_{e2}=2,1$ hyperfine levels in the 5P$_{3/2}$ state. It creates an incoherent transparency window at the expense of the EIT signal. Thus, a narrow EIT signal is not distinctly visible. This is observed  at higher values of the control powers too.
	
	\begin{figure} 
		\centering
		\includegraphics[width=0.65\textwidth]{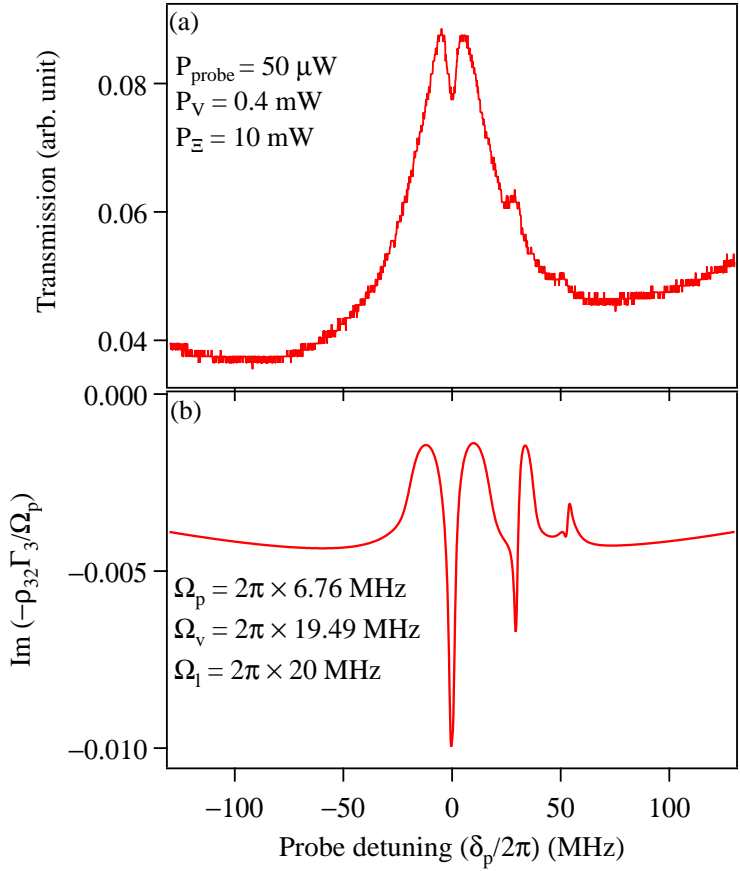}
		\caption{EIA spectrum. (a) Experimental data with probe and control powers as shown. (b) Theoretical result with probe and control Rabi frequencies as shown. The Rabi frequency of the controls does not correspond to the powers shown in part (a), for the reasons explained in the text.}
		\label{fig:EIA}
	\end{figure}
	
	When the two control beams are acting simultaneously, we observe EIA dip over the broader vee EIT. This has been illustrated in Fig.~\ref{fig:EIA}(a). With both the control beams at resonance, the EIA occurs at zero probe detuning. When the vee-control is detuned by about $+25$ MHz from resonance and the ladder-control by about $-25$ MHz, the EIA occurs at the probe detuning of 25 MHz in agreement with the condition we stated earlier. 
	
	The theoretical analysis has been presented earlier \cite{BHN2015} and only the relevant parts are discussed here. We have considered the three hyperfine levels in the 5D$_{5/2}$ manifold with the branching ration of 0.65 \cite{HEA1961}. The theoretical spectrum shown in part (b) of the figure,
	is shown for the EIA condition (that Rabi frequencies of the two control beams are equal) --- the Rabi frequencies for the values of the control powers does not reproduce the experimental data. This is because the theory assumes a uniform intensity for the two control beams. In addition, magnetic sublevels of the hyperfine levels are ignored \cite{GWR2004}.

	For the trace in \ref{fig:EIA}(a), the power in the probe beam is 50 $\upmu$W. The ladder-control power is $10$ mW (corresponding to a Gaussian peak intensity of $838$ mW/cm$^2$) and the vee-control power is $0.4$ mW (corresponding to a Gaussian peak intensity of 34 mW/cm$^2$). Theoretically, to obtain an EIA the Rabi frequencies of the two control beams should be equal. Assuming this condition \cite{FOO2007}, the ratio of the intensities in the ladder- and the vee-control has to be 
	\begin{equation}
		\dfrac{I_{\Xi}}{I_{V}} =\left(\dfrac{\lambda_V}{\lambda_{\Xi}}\right)^3 \dfrac{\tau_{\Xi}}{\tau_V}
	\end{equation}
	 Substituting the values of the wavelengths and the lifetimes of the transitions \cite{SGO2008,STE2015}, we see that this ratio is 9.26. The ratio of intensities at the center of the Gaussian for the result in Fig.~\ref{fig:EIA}(a) is almost three times more. We list a few reasons for this to be so. 
	 \begin{enumerate}
		 \item The beams used in the experiment have different Gaussian profiles and we do not expect them to align within the cell.
		 \item The system under consideration has multiple vee + ladder systems with different relative strengths of the two controls formed because of the Zeeman sublevels, thus making the splitting of EIA apparent only at high control intensties \cite{GWR2004}.
		 \item The absorption peak is not entirely a coherent EIA effect but also has contributions from incoherent phenomena.
	 \end{enumerate}
 
One way to understand the contribution of incoherent effects in EIA will be to consider what effect does the ladder system have on an already present vee EIT. At large detuning of the probe (larger than width of the ladder EIT), the presence of the ladder control does not have any effect qualitatively. Near resonance, when the ladder control is introduced, some additional atoms are driven out of the interacting levels (by optical pumping effects) and a small fraction lie on the two excited states ($F_{e2}=3$ and $F_{e3}=4$). This would decrease the number of atoms in the ground state ($F_g=2$). On one hand the probe transparency is increased just because there are fewer number of atoms to absorb it, but on the other hand probe absorption could increase due to  reduction in probe saturation as more atoms are excited from $F_{e2}=3$ by the ladder control.
Thus, optical pumping and saturation effects, depending on which of the phenomena is dominant, either subtract from or add to the EIA signal. The role of redistribution among the Zeeman sublevels is more complicated. The strong $\pi$-polarized vee-control induces optical pumping such that there is maximum population in the $m_{F_g} = 0$ sublevel, and decreasing as $|m_{F_g}|$ increases. In this configuration, the atoms are less strongly coupled to both the vee-control and the $\sigma^{\pm}$ polarized probe beam on the $F_g=2 \to F_{e2} = 3$ transition. Some of the atoms excited to the $F_{e2} = 3$ level by the probe also undergo a rearrangement due to the strong $\pi$-polarized ladder control. In both cases of $F_{e3} = 4$ and $F_{e3} = 3$ levels, the atoms tend to accumulate on the $m_{F_{e2}} = 0$ sublevel, but differ quantitatively. Decay from this sublevel contributes to populate the $m_{F_g} = 0$ sublevel in the $F_g=2$ hyperfine level as it is most strongly coupled. The other sublevels are less populated due to weaker couplings to corresponding excited sublevel and fewer number of atoms in them. Consequently, the number of atoms strongly coupled to the vee-control decreases. This could thereby contribute to the reduction of the the strength of the vee plus ladder EIA as well.

\begin{figure}
		\centering
		\includegraphics[width=0.65\textwidth]{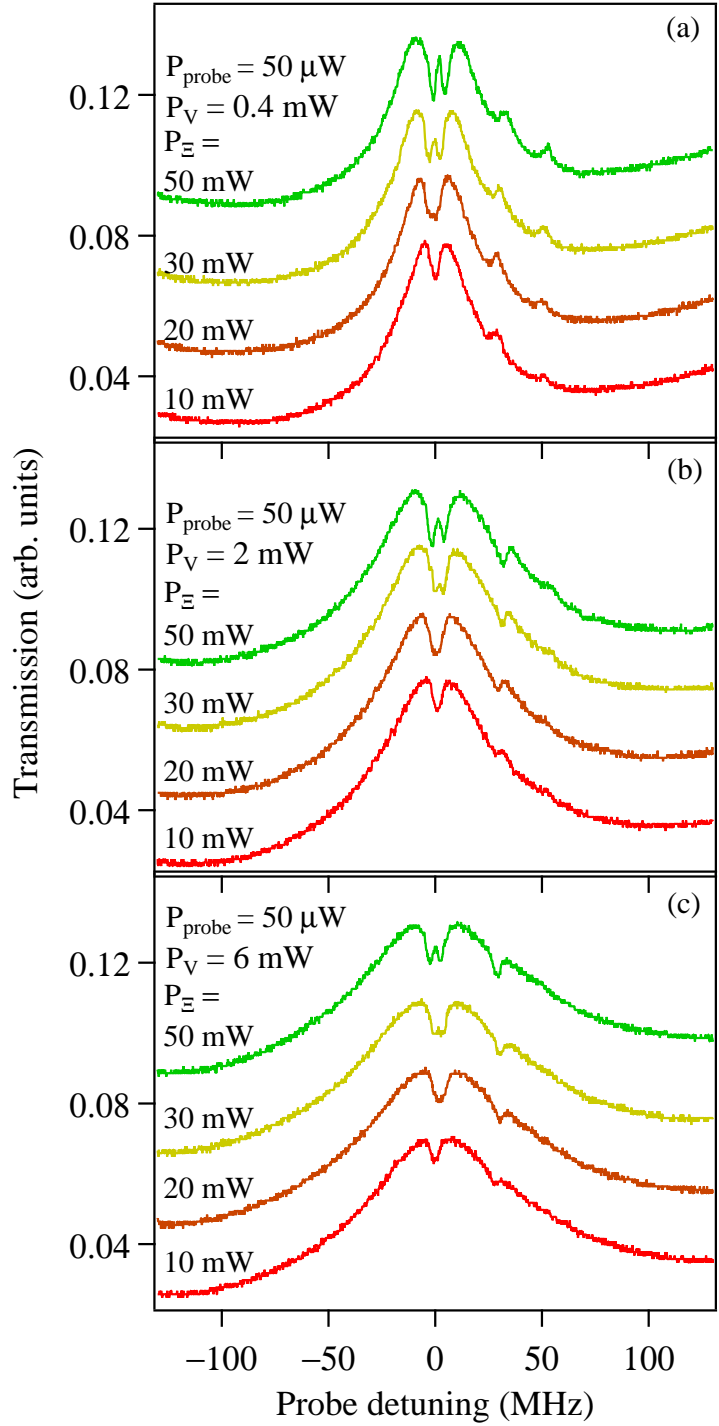}
		\caption{EIA spectra for a fixed value of power in the vee-control beam and varying powers in the ladder-control beam, as listed. (a) $P_V = 0.4 $ mW. (b) $P_V = 2.0 $ mW. (c) $P_V = 6.0 $ mW. The traces are offset for clarity.}
		\label{fig:manyEIAs}
	\end{figure}
	
	We have studied EIA for different powers in the two control beams, as shown in Fig.~\ref{fig:manyEIAs}. Each of the parts shows the spectrum for a fixed power in the vee-control and a varying power in the ladder-control. In Fig.~\ref{fig:manyEIAs}, it can be seen that as the power in the ladder-control is increased, the EIA feature gets broader and eventually splits creating an EIT within. This occurrence of EIT within EIA has been theoretically predicted as well \cite{BHN2015}. The evolution of the relative heights of the EIT and the EIA signal is interesting. At low vee-control powers, it is a significantly strong EIT over an EIA feature. With higher vee-control power, this EIT feature declines in strength and almost vanishes into the EIA as the vee-control power is further increased. This can be attributed to optical pumping of atoms from the interaction levels. 	
	
	\begin{figure}
		\centering
		\includegraphics[width=0.65\textwidth]{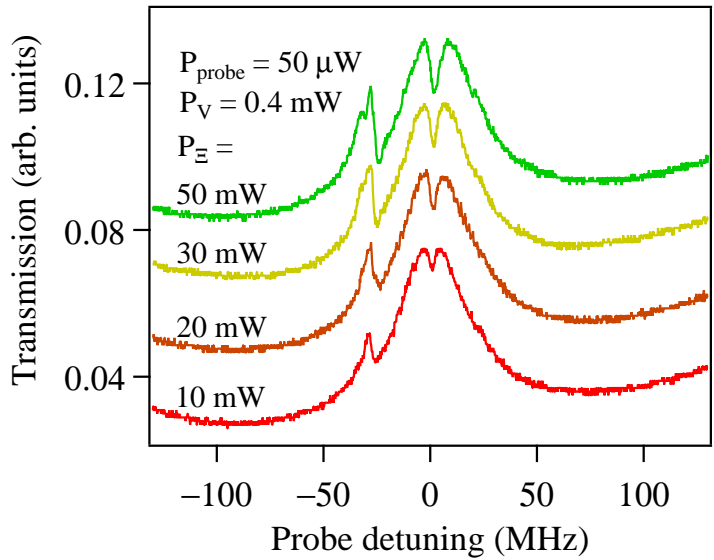}
		\caption{EIT spectra for the $ F_{e3} = 3$ hyperfine level with controls on resonance, for fixed value of power in the vee-control beam and varying powers in the ladder-control beam, as listed. The traces are offset for clarity.}
		\label{fig:EIAF3}
	\end{figure}
	
	We observe that the EIA dip, for $F_{e3} = 3$ at resonance shown in Fig.~\ref{fig:EIAF3}, does not split for any power of the ladder-control. Here, optical pumping to other hyperfine levels is quite high because of several available decay channels in the ladder part of the system.	The EIA feature remains intact but only increases in height due to the combined effect of the coherent and incoherent phenomena, discussed earlier. It is noteworthy that the width of the EIA is smaller than the corresponding EIT in Fig.~\ref{fig:EIT}(b), indicating the coherent origin of the EIA dip.

	\section{Conclusion}
	In this paper, we have reported the experimental observation of electromagnetically induced absorption in a vee + ladder system. We have also studied the effects of the strengths of the two control beams on the EIA lineshape. A theoretical analysis has shown that the appearance of an absorption dip has contributions from several incoherent phenomena besides EIA, in real experimental systems. The system considered exhibits a conversion between EIT and EIA, thus enabling the tunability of the light from sub-luminal to super-luminal transmission. An experimental observation of a vee + ladder EIA is the first step towards realizing the possibilities mentioned in the introduction. 
	
	\section*{Acknowledgement}
	We would like to thank Kavita Yadav for insightful discussions on the interpretation of the results. We would also like to thank Anish Bhattarcharya for help with the experiment. This work was supported by the Department of Science and Technology, India.

\end{document}